\documentclass[%
 reprint,
superscriptaddress,
 amsmath,amssymb,
 aps,
 prl,
]{revtex4-2}
\usepackage{graphicx}
\usepackage{dcolumn}
\usepackage{bm}
\usepackage{color}

\newcommand{\vect}[1]{\boldsymbol{#1}}
\usepackage{lineno}

\begin{document}

\title{Ligand-induced protein dynamics differences correlate with protein--ligand binding affinities: An unsupervised deep learning approach}
\author{Ikki Yasuda}
\affiliation{Department of Mechanical Engineering, Keio University, Yokohama, Kanagawa 223-8522, Japan}
\author{Katsuhiro Endo}   
\affiliation{Department of Mechanical Engineering, Keio University, Yokohama, Kanagawa 223-8522, Japan}
\author{Eiji Yamamoto}   
\affiliation{Department of System Design Engineering, Keio University, Yokohama, Kanagawa 223-8522, Japan}
\author{Yoshinori Hirano}   
\affiliation{Department of Mechanical Engineering, Keio University, Yokohama, Kanagawa 223-8522, Japan}
\affiliation{Laboratory for Computational Molecular
Design, RIKEN Center for Biosystems Dynamics Research
(BDR), Suita, Osaka 565-0874, Japan}
\author{Kenji Yasuoka}   
\email{yasuoka@mech.keio.ac.jp}
\affiliation{Department of Mechanical Engineering, Keio University, Yokohama, Kanagawa 223-8522, Japan}

\begin{abstract}
Prediction of protein--ligand binding affinity is a major goal in drug discovery. Generally, free energy gap is calculated between two states (e.g., ligand binding and unbinding). The energy gap implicitly includes the effects of changes in protein dynamics induced by the binding ligand. However, the relationship between protein dynamics and binding affinity remains unclear. Here, we propose a novel method that represents protein behavioral change upon ligand binding with a simple feature that can be used to predict protein--ligand affinity. From unbiased molecular simulation data, an unsupervised deep learning method measures the differences in protein dynamics at a ligand-binding site depending on the bound ligands. A dimension-reduction method extracts a dynamic feature that is strongly correlated to the binding affinities. Moreover, the residues that play important roles in protein--ligand interactions are specified based on their contribution to the differences. These results indicate the potential for dynamics-based drug discovery.
\end{abstract}

\maketitle

\section*{Introduction}
In computational drug discovery, the estimation of binding affinities between the target proteins and ligands is one of the main goals. Various approaches have been proposed and performed for both physics-based and data-driven methods. In physics-based approaches, protein--ligand free energy calculations have been widely conducted using free energy perturbation and thermal integration methods, and the results agree well with experimental data \cite{aldeghi2016accurate, cournia2017relative, gapsys2020large}.
However, despite the high accuracy, the high-calculation cost has prevented its practical use \cite{abel2017advancing}.
Data-driven approaches, such as scoring functions for docking, quantitative structure--activity relationship method with machine learning, and deep learning methods have been studied over the past few decades \cite{zhang2017machine, shen2020machine}.
Deep learning approaches can grasp important characteristics automatically from the high-dimensional data of proteins and ligands. The approaches for protein--ligand affinity prediction have succeeded in finding relevant patterns in 3D structures \cite{gomes2017atomic, jimenez2018k,cang2018representability,stepniewska2018development} and protein and ligand sequences \cite{ozturk2018deepdta,karimi2019deepaffinity} using supervised learning with a sufficient amount of dataset. Although there are widely used databases such as PDB-bind \cite{wang2004pdbbind} and DUD-E \cite{huang2006benchmarking} for protein--ligand-binding data, an efficient approach cannot be determined if the available dataset is limited \cite{shi2021deep}. 

Protein dynamics play an important role in biological phenomena. All-atom molecular dynamics (MD) simulations are powerful tools that can generate a large amount of dynamic data and analyze protein dynamics at the atomic level, along with experimentation \cite{fernandez2016unravelling} and course-grained computations \cite{bahar2010global,bahar2010normal}. MD data analysis for protein dynamics  has focused on protein fluctuations, relaxation time, stability, and state transitions. The commonly used methods are root mean square deviation, principal component analysis, relaxation time analysis, decomposition cross-correlation maps, and root mean square fluctuation \cite{yang2020larmd, jin2019communication, yamamoto2021universal,mitsutake2011relaxation}. For protein and ligand systems, these methods have revealed that proteins undergo dynamic changes before and after ligand binding \cite{komatsu2020drug, stanley2016pathway}. In addition to traditional methods, machine learning approaches have recently been proposed to obtain useful information from MD trajectories of protein systems \cite{noe2020machine}. For the description of folding--unfolding events, VAMPnets \cite{mardt2018vampnets} could replace handcrafted procedures for creating a Markov state with deep neural networks (DNNs). Moreover, an autoencoder that can detect allosteric dynamics from the time fluctuation of a protein has been proposed \cite{tsuchiya2019autoencoder}.

Although these methods could identify ligand-induced dynamics changes in proteins, integration of the information and representation of the dynamics with a few variable is still challenging, consequently making it difficult to directly link the dynamics properties with ligand affinity. In this study, we propose a novel method to predict binding energies from proteins’ dynamics change upon ligand binding, based on a deep learning approach for MD data analysis \cite{endo2019detection}. In contrast to general approaches for MD trajectory-based supervised machine learning \cite{ferraro2020machine, riniker2017molecular}, our method uses raw MD trajectories of a ligand binding site with different kinds of ligands, and quantitatively measures differences in the dynamics using unsupervised learning. Overall, the workflow consists of simplification of MD trajectories as local dynamics ensemble (LDE), calculation of dynamics difference as Wasserstein distance, and extraction of variables and detection of the contributing residues.  We verified the method in two systems, bromodomain 4 (BRD4) \cite{fujisawa2017functions,cochran2019bromodomains} and protein tyrosine phosphatase 1B (PTP1B) \cite{johnson2002protein,verma2017protein} systems, which have used for benchmark of free energy calculation in previous studies \cite{wang2015accurate, song2019using, he2020fast,aldeghi2016accurate,gapsys2020large}. We illustrate a strong correlation between the extracted feature and binding energies, and suggest that the feature relevant to dynamics can work as a predictor of binding energy. In addition, the dynamic differences are interpreted from potential bindings between specific residues and ligands. We believe this approach provides a unique perspective of ligand-induced dynamics of protein towards a further understanding of ligand functions and binding mechanism.  

\section*{\label{sec:result}Results}
\subsection*{Determination of LDE of protein--ligand systems }
An overview of our method is shown in Fig. 1. First, we performed a total of 6.6-$\mu$s all-atom MD simulations [10 ligand-bound (holo) protein and one ligand-unbound (apo) protein system] to obtain trajectories of the BRD4 systems. To exclude the dependency of the initial conditions, three 200-ns independent production runs were executed with different initial velocities for each system.

To grasp the features of protein dynamics from MD simulation data within a moderate data size for the DNN, the concept of LDE \cite{endo2019detection} was employed. LDE should have an appropriate selection of particles and time length to contain important dynamics of interest. We assume that the behavior of amino acids is sensitive to the presence of a ligand, that is, the binding site shows representative dynamics that are induced by ligand interactions and is suitable target for LDE. The selection of binding sites is detailed in the Methods section. Furthermore, to reduce noise and computation cost, the center of mass of the heavy atoms of each residue in a ligand-binding site were used as the trajectories of LDE. Notably, this selection includes no information on the bound ligand, making it possible to directly compare the behavior of the ligand-binding sites between systems. As for the time length, we selected 100-ps order dynamics to capture the side-chain movements \cite{bahar2010normal}, which was sufficient to identify differences in local dynamics (Fig. S1). 

The properties of the LDE with the selection above are presented in the supplementary material. The value of $g(\vect{x})$ used as characteristics of the dynamics were distributed similarly regardless of the initial conditions (Fig. S2). Moreover, the local dynamics were distributed evenly throughout the MD simulations, suggesting the fact that they were not influenced by the slowest fluctuations (Fig. S3). These results indicated that the local dynamics were robust in terms of the initial conditions and long-term dynamics.

\begin{figure*}
\includegraphics[width=170 mm, bb=0 0 504 302]{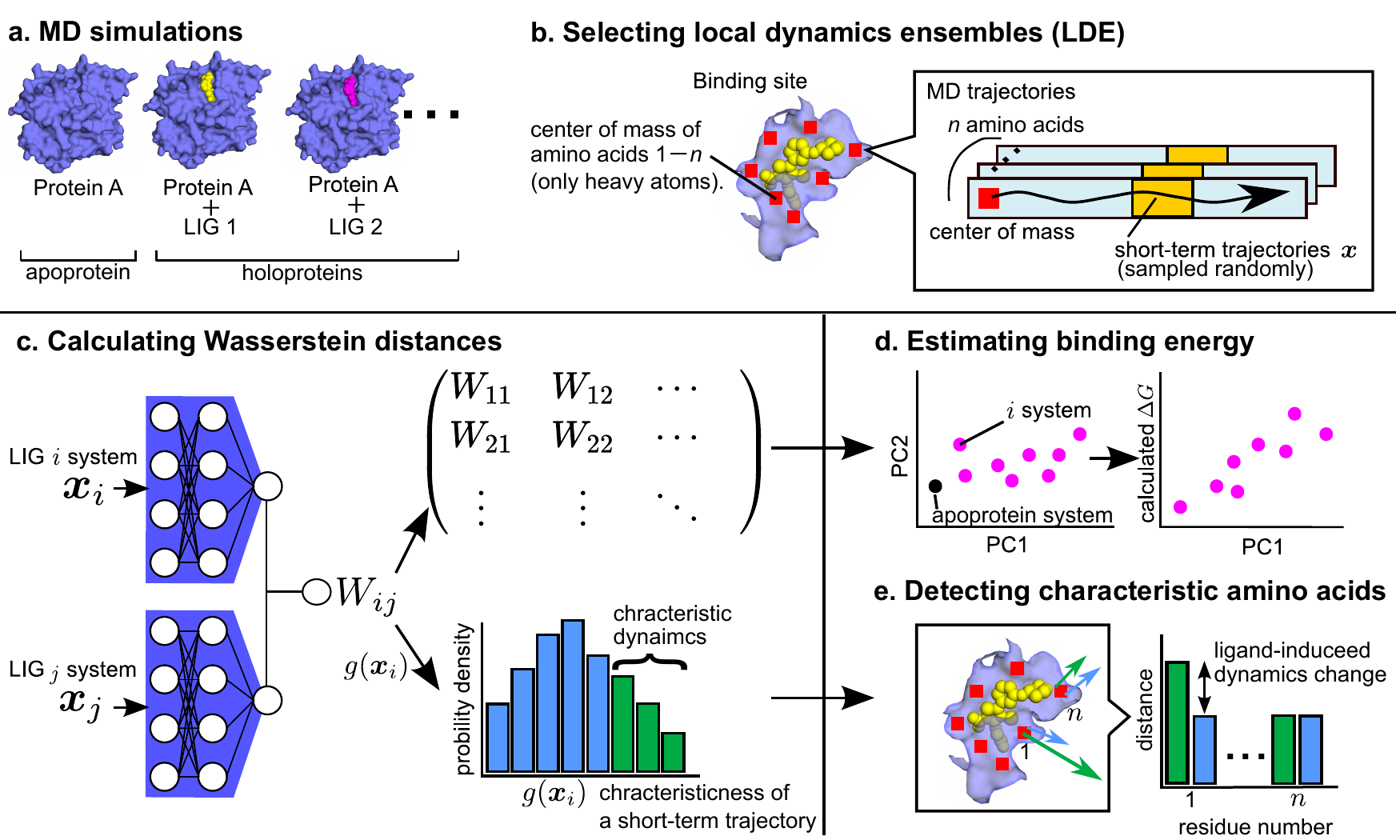}
\caption{\label{fig:fig1}The workflow to detect differences in protein dynamics of systems with different bound ligands. (a) Molecular dynamics (MD) simulations for protein--ligand complexes and apoprotein systems. (b) Determination of the local dynamics ensemble (LDE). (c) Calculation of Wasserstein distances using deep neural networks (DNNs). Wasserstein distances are calculated for all pairs of system. The output of DNN for each short-term trajectory  in system $i$ is denoted as $g(\vect{x}_i)$. The short-term trajectories expressed in green are characteristic of system $i$, whereas those in blue are similar to system $j$. (d) Embedding of the Wasserstein distance into the lower dimensions and comparing the first principal component with the binding energies. (e) Detection of amino acids whose dynamics are changed by ligand interactions. A trajectory of LDE which consists of short-term trajectories of multiple residues are decomposed by residues, and their moving distance are compared.}
\end{figure*}

\begin{figure*}
\includegraphics[width=170 mm, bb=0 0 482 482]{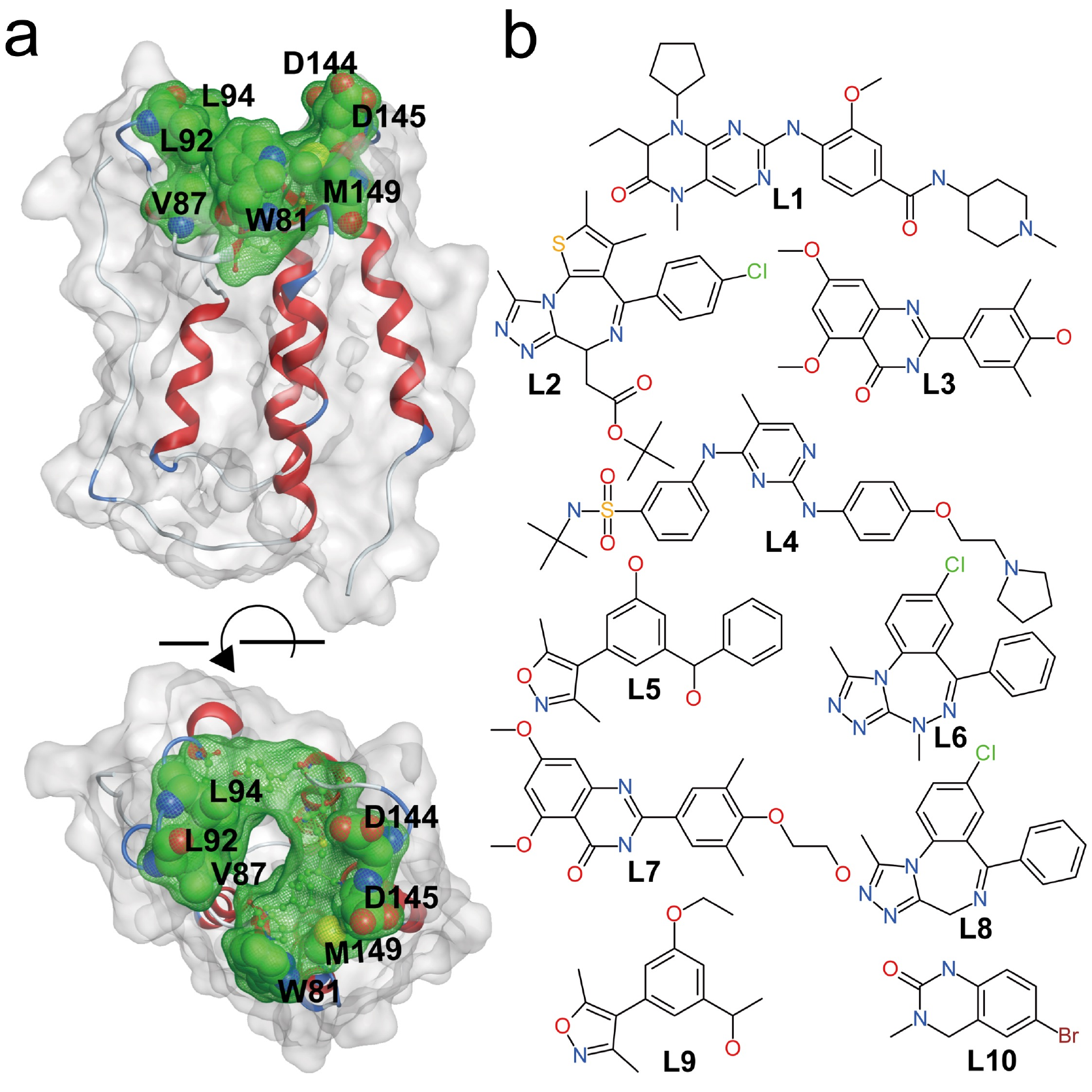}
\caption{\label{fig:fig2} The bromodomain 4 (BRD4) system. (a) Ribbon diagram of the configuration of BRD4 (Protein Data Bank ID: 2OSS) superimposed on the molecular surface. The ligand-binding site, residues in the ligand-binding site, and key residues are shown in green meshed molecular surfaces, green balls and sticks, and green space-filling with labels, respectively. (b) Chemical structures of ligands L1--L10. The L1 to L10 labels correspond to the ligands of the 1 to 10 holoprotein systems.}
\end{figure*}

\subsection*{\label{sec:result2}Calculation of Wasserstein distances using a DNN}
To measure differences in LDE quantitatively, Wasserstein distances \cite{villani2009optimal,arjovsky2017wasserstein} were calculated using two identical lobes of a DNN. The input data size for the DNN was $(n,\Delta,d)$, where $n,\Delta,d$ represent the number of amino acids in the ligand-binding site, time steps of the LDE, and dimension, respectively. The time series of displacement vectors of the amino acids from a time step, $\bm{x}(t_0+\Delta)-\bm{x}(t_0)$, were used as the input, where $\bm{x}(t_0)$ is the LDE trajectory data, and in the training of the DNN, $t_0$ was randomly sampled from the total MD simulation time. Rotation and translation were removed from the raw trajectories by fitting the configurations to an identical structure in the backbone atoms of the selected residues. The displacement considers the diffusion of residues in the LDE time without the average structure and slow fluctuations. Wasserstein distances were calculated for all pairs of $N$ ligand systems, resulting in a distance matrix of $N$ $\times$ $N$. It is evident that the apoprotein system $\rm{S_0}$ shows a relatively larger distance from the holoprotein system than that between the holoprotein and another holoprotein system [Fig. 3(a)].  

\subsection*{\label{sec:result3}Extraction of a variable for protein dynamics}
The high dimension of the distance matrix makes it difficult to understand the whole differences in systems and extract simple features. Therefore, the distance matrix was embedded into $N$ vectors in a low-dimensional space that represented the systems. Then, the first and second principal components were extracted. 

The distance embedding of LDE demonstrated clear differences in protein's short-term dynamics in the systems [Fig. 3(b)]. As shown in the distance matrix, the apoprotein system $\rm{S_0}$ is separated from the holoproteins. Moreover, systems with lower-affinity ligands tend to position near apoprotein compared to systems with higher-affinity ligands. The link between the first principal component (PC1) and binding affinities was quantitatively evaluated by comparing it to the binding energies calculated in a previous study \cite{aldeghi2016accurate}. Pearson's product moment correlation coefficient ($r$) between PC1 and the binding energies was 0.82 [Fig. 3(c)]. 

\begin{figure*}
\includegraphics[width=170 mm, bb=0 0 504 367]{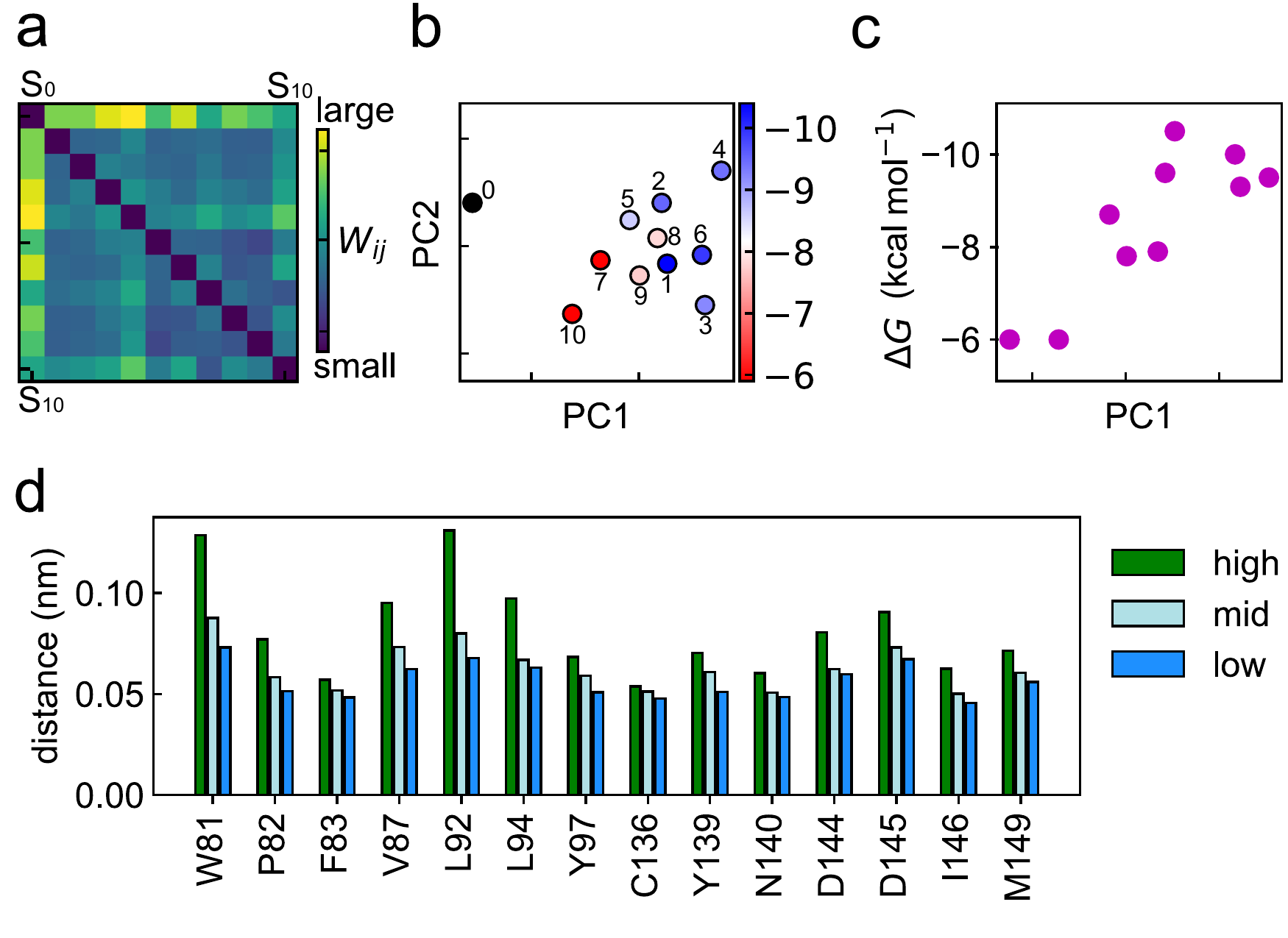}
\caption{\label{fig:fig3} Detection of differences in the protein dynamics of the BRD4 system. (a) Distance matrix calculated using the LDE distributions. A large Wasserstein distance corresponds to a large difference in the protein dynamics. (b) Embedding of the distance matrix. The first principal component (PC1) and the second principal component (PC2) from the three-dimension vectors are shown. Each circle corresponds to a system. Points are colored according to the binding energies, and the apoprotein system is in black. The binding energies calculated in a previous study were used \cite{aldeghi2016accurate}. (c) Correlation between PC1 and the binding energies. (d) Moving distance of the center of mass from a reference point in short-term trajectories of the apoprotein system. System 4 was used as the reference for comparison. Short-term trajectories were divided into three groups based on $g(\vect{x})$. High and low groups consisted of short-term trajectories with top and bottom 5\% $g(\vect{x})$ values, respectively. The high group had characteristic trajectories of apoprotein , whereas the low group had similar trajectories to holoprotein. The distances were averaged over the latter half of the LDE trajectories.}
\end{figure*}

\subsection*{\label{sec:result4}Contribution of amino acids to dynamic differences}
Given that the features of dynamic differences could be an indicator of ligand affinity, specification of highly influenced amino acids in their dynamics is useful for understanding the binding mechanism. To analyze the trajectories in one system, the $g(\vect{x})$ was used. When comparing a pair of systems, the function $g(\vect{x})$ can distinguish characteristic trajectories in a system from those similar to the other system.   

To detect important residues, the amino acids that had different dynamics in the apoprotein system from a holoprotein system (system 4) were predicted using $g(\vect{x})$. In the embedding figure, system 4 was the most distant from the apoprotein system in PC1. First, $g(\vect{x})$ for each short-term trajectory in the apoprotein system to system 4 was calculated. We note that one short-term trajectory of LDE contains the dynamics of multiple residues in the same time frames. Therefore, even if the two short-term trajectories have the same value, the attributable residue might be different. Second, MD trajectories in the apoprotein system were grouped into three categories according to the value of $g(\vect{x})$, with the top 5\% of $g(\vect{x})$ (high), bottom 5\% of $g(\vect{x})$ (low), and others (mid). Finally, the average distances that the center of mass of each amino acid moved were calculated for each group. 

The characteristic behavior in the apoprotein system demonstrated large movement in all amino acids [Fig. 3(d)], indicating that the apoprotein was more flexible than the holoprotein. In particular, the residues Trp81, Val87, Leu92, Leu94, Asp144, and Asp145 showed relatively large gaps between the high and low groups. In other words, this shows that the behavior in system 4  was more constrained, which was probably because of ligand interactions. Thus, these residues may play an important role in protein--ligand binding. The mid groups were calculated to demonstrate that the value of $g(\vect{x})$ corresponds to the distances moved by the residues. Otherwise, $g(\vect{x})$ may reflects properties such as the direction of motion.   

\begin{figure*}
\includegraphics[width=175 mm, bb=0 0 504 302]{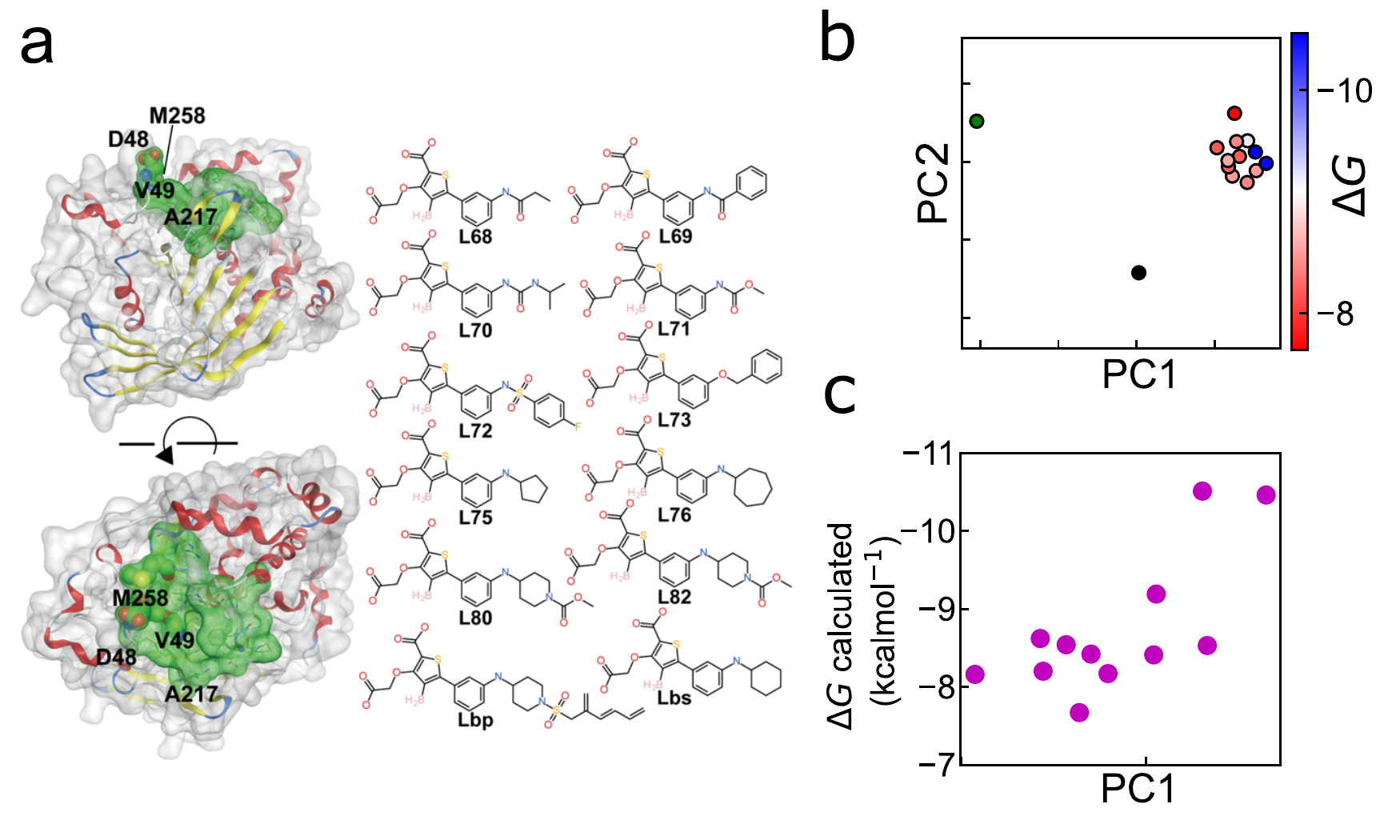}
\caption{\label{fig:fig4} Analysis of ligand-induced dynamics in the phosphatase 1 B (PTP1B) systems. (a) Ribbon diagram of the configuration of PTP1B (Protein Data Bank ID: 1OEM) superimposed on the molecular surface. The ligand-binding site, residues in the ligand-binding site, and key residues are shown in green meshed molecular surfaces, green balls and sticks, and green space-filling with labels, respectively. The labeled residues were selected in a similar manner to BRD4 using $g(\vect{x})$. (right column) Chemical structure of ligands. The binding structures are shown in Fig. S7. (b) Embedding of the distance matrix. The first principal component (PC1) and the second principal component (PC2) from the three-dimension vectors are shown. Each circle corresponds to a system. Points are colored based on binding energies. The crystal apoprotein system is colored in green, whereas the pseudo apoprotein system is in black. The binding energies calculated in a previous study were used \cite{he2020fast}. (c) Correlation between PC1 and binding energies.} 
\end{figure*}

\section*{\label{sec:discussion}Discussion}
In this study, we presented a deep learning approach that can determine the differences in protein behavior associated with the binding of different ligands. MD simulation data of apo and holoprotein systems were reduced to short-term trajectories of the LDE. Pairwise Wasserstein distances were calculated using the DNN. Finally, the variables were extracted using dimension-reduction methods for comparison with binding energies, and the characteristic short-term trajectories in a system were determined using $g(\vect{x})$ for the detection of key residues.      

For the BRD4 systems, there was a strong correlation between PC1 and binding energies. Here, we also investigated another protein system, tyrosine phosphatase 1 B (PTP1B), in order to assess the potential universality of our method. To obtain trajectories, we performed all-atom MD simulations of 14 systems of PTP1B [twelve holo-PTP1B with different ligands and two apo-PTP1B systems (Fig. 4(a))]. For apo-PTP1B, we prepared two initial structures: the original apo-PTP1B, which was the X-ray crystallographic structure of the ligand-unbound form and the pseudo apo-PTP1B, in which the ligand was removed from the ligand-bound form. Following MD simulations, the MD data were analyzed in a manner similar to the analysis of BRD4. The differences in the systems are illustrated in Fig. 4(b). Embedding could identify apoprotein systems that are quite different from the corresponding holoprotein systems. Furthermore, PC1 showed a strong correlation with binding energies, with Pearson's $r = 0.72$. It should be noted that all ligands in the prepared PTP1B systems had a relatively high affinity, which might have made it difficult for dynamic properties to accurately correspond to binding energies. Complexes with a wider range of affinities would be more desirable. Nonetheless, the calculated dynamic property tended to reflect the affinities of the ligands.

Our method also detected the amino acids that were influenced by the ligand, indicating that they may play important roles in ligand binding. Comparison of apo-BRD4 dynamics with that of BRD4 bound to ligand 4 indicated that the largest dynamic change occurred at Trp81, Val87, Leu92, Leu94, Asp144, and Asp145. The crystal structures showed that these amino acids were located around the loop domain and were exposed to solvation (Fig. S4). Combining both observations, we assume that these residues freely move in an apoprotein, and the binding of the ligand restricts the movements by making some interactions. Interestingly, previous studies have indicated that hydrophobic interactions are likely to be created with these residues when some types of ligands are bound \cite{ran2015insight,wang2019theoretical}. In addition, the hot spots predicted in those studies largely agreed with the sum of the dynamics-influenced residues detected and the directly binding residues in the crystal structure of system 4 (Pro82, Asn140, and Ile146). Therefore, we could specify some important residues based on the ligand-induced dynamics at the residue level. 

The main hyperparameters in our method are involved in MD simulations and LDE selection. First, MD simulations are required to be sufficiently long to sample the LDE. The number of data points exponentially reduced the error in the Wasserstein distances (Fig. S5). Second, the LDE time length should be selected so that the resulting feature corresponds to a property of interest. In the case of binding affinity, the appropriate length was suggested to be that for the side-chain movement. In Fig. S6, the results from different LDE time are compared, and the robustness to the selection is shown. Finally, the selection of binding sites can be addressed by repeating the process multiple times. In the initial analysis, the input residues can be determined based on the distances to the ligand atoms. The proposed method can detect potentially important residues. In subsequent run, the extracted residues can be used to obtain a feature of the important protein dynamics. 

A point for further improvement is in the identification of ligand-influenced residues from $g(\bm{x})$. Characteristic dynamics are detected by $g(\bm{x})$, and we need to determine the appropriate measurements for the characteristics. In the BRD4 systems, the characteristic behavior largely corresponded to the distances. However, the characteristic behavior might be more explicit if other criteria are used, such as the direction of movement. 

In conclusion, this study suggests that differences in ligand-induced dynamics at a binding site could be an indicator of protein--ligand binding energies. While our test cases were relatively rigid proteins that are widely used as benchmarks for free energy calculation, flexible proteins are interesting targets for further investigation. Dynamic properties may play a more important role in these systems. It has been reported that for a type of flexible protein, ligand-induced flexibility contributes to entropy gain in ligand binding, thus leading to higher affinity and longer residence of the ligands \cite{amaral2017protein}. Furthermore, a similar approach could be used to study other protein--ligand binding events. For instance, it is interesting to evaluate the relationship between allosteric dynamics and ligand function, as has been done in a few previous studies \cite{jin2019communication, ferraro2020machine}. Another potential application would be predicting the effects of protein mutations from the dynamics of the ligand. In this case, the relationship between the protein and ligand would be analyzed in a manner opposite to that employed in the present case; that is, identical particles of the ligand would be used for LDE, while the protein molecules would vary slightly because of the mutation. We believe that the understanding of protein dynamics using ligand interactions will provide deeper insight into the function of ligands, and dynamics-based approaches would contribute to further developments in computational drug discovery.

\section*{\label{sec:method1}Method}
\subsection*{\label{sec:method1}System setup and MD simulations}
For the BRD4 systems, the initial structures and topologies for proteins and ligands were considered according to a previous study \cite{aldeghi2016accurate}. The protein structures with and without ligands were solvated in a TIP3P \cite{jorgensen1983comparison} cubic box with a minimum distance of 1.0 nm. The systems were neutralized by adding $\rm{Na^+}$ or $\rm{Cl^-}$ ions.

All-atom MD simulations of the systems were performed using GROMACS 2019.6 \cite{abraham2015gromacs}. The particle mesh Ewald method \cite{hess1997lincs} was used to evaluate the electrostatic interactions with a cut-off radius of 1.2 nm, and van der Waals interactions were switched between 1.0 and 1.2 nm. The bonds with H atoms were constrained with LINCS \cite{hess1997lincs} in the order of 4. For the prepared systems, energy minimization was carried out until the maximum force reduced to less than 10.0 kJ/mol, using the steepest descent method. Then restrained MD simulation was performed in a $NVT$ ensemble for 100 ps and subsequently in an $NPT$ constant simulation for 100 ps. During both equilibration processes, position restraints were executed on the heavy atoms of the ligand and protein atoms. The temperature and pressure were regulated using the velocity-rescaling method \cite{bussi2007canonical} and Berendsen methods \cite{gros1990inclusion}, respectively. Finally, three individual production runs were performed for 200 ns in the $NPT$ ensemble for each system with a random initial velocity generated to simulate different initial conditions. In the production runs, pressure was controled with Parrinello-Rahman pressure coupling method \cite{parrinello1981polymorphic}. The trajectories were recorded every 2 ps. 

For PTP1B systems, 12 complexes and two types of apoprotein systems were prepared for MD simulations. Ten complexes and pseudo-apoprotein systems were constructed from the initial structures of PTP1B and the ligand used in the previous study by Song et. al \cite{song2019using}. In addition, another apoprotein was modeled from the crystal structure of the PTP1B structure with no ligand (PDB ID: 1OEM) by homology modeling. Missing atoms were complemented using MOE \cite{chemical2016molecular}, where we selected a structure without the ${\alpha}$-helix. To distinguish between the two apo-protein systems, we denoted the apoprotein from the study by Song et. al as a pseudo-apo, which is originally complex and no ligand was added in our study. We called the apoprotein created from the crystal structure of the apoprotein as a crystal apoprotein. Proteins and water were parameterized by Amber ff14SB \cite{maier2015ff14sb} and TIP3P \cite{jorgensen1983comparison}, respectively. The parameters for the ligand were generated using GAFF\cite{wang2006automatic} and parameter files in the study conducted by Song et. al. Subsequently, the systems were solvated into a cubic box, with the thickness of the water shell set to 1.2 nm. Next, the system was neutralized with $\rm{Na^+}$.

MD simulations for PTP1B systems were performed similarly to the BRD4 systems, except for a few points. Energy minimization was performed for 10,000 steps, and the equilibration process was continued for 200 ps in both the \emph{NVT} and \emph{NPT} ensembles. Considering stability, we removed the first 50 ns of the trajectories in the following deep learning analysis in PTP1B systems.

\subsection*{Selection of the LDE}
In BRD4, 14 residues in the binding site were determined based on the activity ratio in MD simulations and the previous work by Aldeghi et. al \cite{aldeghi2016accurate}. To obtain the activity ratio for each residue, the distances between the heavy atoms of the protein and those of the ligand were calculated. We defined the activity ratio as $N_{contact}/N_{total}$, where $N_{contact}$ is the number of frames in which the residues are in contact with the ligand, and $N_{total}$ is the total number of output frames of the MD simulation. The protein--ligand contact was confirmed if any heavy atom of the residue was positioned within a distance $r$ from any of the ligand heavy atoms, where $r$ was set to 0.5 nm. Residues that showed an activity ratio of more than 0.5 were regarded as binding residues. Activity ratio was calculated for the three simulations for each system, and the sum of the satisfactory residues were used. In addition to the distance-based criteria, the selection of binding site residues by Aldeghi et. al was considered to further limit the number of binding site residues for input. All residues from the selection of Aldeghi et. al satisfied the distance-based criteria. Based on these two criteria, 14 residues were selected (Trp81, Pro82, Phe83, Val87, Leu92, Leu94, Tyr97, Cyc136, Tyr139, Asn140, Asp144, Asp145, Ile146, and Met149).

For PTP1B systems, the binding site residues were determined in a manner similar to that for the BRD4 systems. The values of the contact distance and activity ratio were identical to those of BRD4 systems, and we referred to the previous study conducted by Liu et al. \cite{liu2014investigating}. Nineteen residues were chosen as binding sites (Tyr46, Asp48, Val49, Lys120, Pro180, Asp181, Phe182, Gly183, Cys215, Ser216, Ala217, Ile219, Gly220, Arg221, Arg254, Met258, Gly259, Gln262, and Gln266).

In both cases, the time length of the LDE was set to 128 ps (= 2 ps $\times$ 64 steps), which presumably corresponded to the scale of the side-chain movement. 

\subsection*{\label{sec:method3}Wasserstein distance between systems and its calculation using DNN}
The Wasserstein distance between two LDEs is expressed as  
\begin{equation}
    W_{ij}= \sup_{||f||_{L \leq 1} } \mathbb{E}_{\vect{x} \sim \vect{y_i} }\left[ f(\vect{x})\right] - \mathbb{E}_{\vect{x} \sim \vect{y_j}}  \left[f(\vect{x})\right] \label{eq:wd1}
\end{equation}
where $i,j$ are the indexes for the systems, supremum is over all the 1-Lipschitz function $f$, $\vect{x}$ is the short-term trajectory of the LDE, and $\vect{y_i}$ is the LDE of system $i$. The function $f$ is expressed by the DNN, and using the trained function $f^*$, $W_{ij}$ is described as  
\begin{equation}
    W_{ij}= \mathbb{E}_{\vect{x} \sim \vect{y_i} }\left[ f^*(\vect{x})\right] - \mathbb{E}_{\vect{x} \sim \vect{y_j}}  \left[f^*(\vect{x})\right]. \label{eq:wd2}
\end{equation}
The architecture of the DNN is almost identical to that of the previous study by Endo et. al \cite{endo2019detection}, except for two changes. First, the input size was changed to $n \times 64 \times d$, where $n$ and $d$ are the number of amino acids in the LDE and dimension, respectively. Second, the number of nodes was increased to 2048 to adapt to the complexity of the protein systems. 

During the training, LDE trajectories were randomly obtained from the three simulations for each system.

\subsection*{\label{sec:method4}Embedding of Wasserstein distances}
A Wasserstein distance matrix was embedded into vectors in low-dimensional space to satisfy the following equation:
\newcommand{\argmin}{\mathop{\rm arg~min}\limits}
\begin{equation}
    \vect{p}_0,\vect{p}_1,...,\vect{p}_n= \argmin_{\vect{p}_0,\vect{p}_1,...,\vect{p}_n} \sum_{i,j}(W_{i,j}-||\vect{p}_i-\vect{p}_j||)^2 \label{eq:emb}
\end{equation}
where $\vect{p}_i$ is an $n$-dimensional vector that corresponds to system $i$. In this study, the number of dimensions was set to three. The embedded vectors were optimized using simulated annealing and gradient descent. Simulated annealing was employed to explore the global minimum and gradient descent for fast convergence. 

\subsection*{A function $g(\vect{x})$}
A function $g(\vect{x}_i$) \cite{endo2019detection} represents the contribution of one trajectory fraction to the overall differences between two systems. The definition for a LDE trajectory of system $i$ with the referenced system $j$ is expressed as 

\begin{equation}
        g({\vect{x}_i})= \mathbb{E}_{\vect{x}' \sim \vect{y_j} }\left[ f^*(\vect{x}_i) - f^*(\vect{x}')\right].\label{eq:gx}
\end{equation}
This is equivalent to 
\begin{equation}
    W_{ij}= \mathbb{E}_{\vect{x}' \sim \vect{y_j} }\left[ g(\vect{x}_i')\right]. \label{w-gx}
\end{equation}
The calculation was performed using the output of DNN when the input are one short-term trajectory of one system $i$ and the average local dynamics of the other system $j$ in a pair. The value of $g(\vect{x})$ could be used to evaluate one short-term trajectory from a system in the perspective of how much similar they are to the overall dynamics of the other system. For instance, if a short-term trajectory in system $i$ has a small $g(\vect{x})$ when system $j$ is referenced, the short-term trajectory of system $i$ is similar to the average molecular behavior seen in system $j$ and vice versa. $g(\vect{x})$ was calculated every 64 steps. 

\section*{Acknowledgements}

\section*{Author contributions}
K.E., E.Y., and K.Y. conceptualized the research. I.Y. and Y.H. performed the simulations. I.Y and K.E. analyzed the data. I.Y., K.E., E.Y., Y.H. and K.Y. wrote and edited the manuscript. 

\providecommand{\noopsort}[1]{}\providecommand{\singleletter}[1]{#1}%

\end{document}